\documentclass[prb,twocolumn,superscriptaddress,showpacs]{revtex4-1}
\usepackage{graphicx}
\usepackage{siunitx}
\usepackage{bm}
\usepackage{color}
\usepackage{amsmath}
\usepackage{amssymb}
\usepackage{multirow}

\newcommand{\scwo}{Sr$_2$CuWO$_6$}

\begin{document}

\title{Spin wave excitations in the tetragonal double perovskite Sr$_2$CuWO$_6$}

\author{H. C. Walker}
\affiliation{ISIS Neutron and Muon Source, Rutherford Appleton Laboratory, Chilton, Didcot, OX11 0QX, United Kingdom}
\author{O. Mustonen}
\affiliation{Department of Chemistry, Aalto University, FI-00076 Aalto, Finland}
\author{S. Vasala}
\affiliation{Department of Chemistry, Aalto University, FI-00076 Aalto, Finland}
\affiliation{Laboratory of Photonics and Interfaces, \'{E}cole Polytechnique F\'{e}d\'{e}rale de Lausanne, CH-1015 Lausanne, Switzerland}
\author{D. J. Voneshen}
\affiliation{ISIS Neutron and Muon Source, Rutherford Appleton Laboratory, Chilton, Didcot, OX11 0QX, United Kingdom}
\author{M. D. Le}
\affiliation{ISIS Neutron and Muon Source, Rutherford Appleton Laboratory, Chilton, Didcot, OX11 0QX, United Kingdom}
\author{D. T. Adroja}
\affiliation{ISIS Neutron and Muon Source, Rutherford Appleton Laboratory, Chilton, Didcot, OX11 0QX, United Kingdom}
\affiliation{Highly Correlated Matter Research Group, Physics Department, University of Johannesburg, P.O. Box 524, Auckland Park 2006, South Africa}
\author{M. Karppinen}
\affiliation{Department of Chemistry, Aalto University, FI-00076 Aalto, Finland}

\date{\today}
\begin{abstract}
\scwo\ is a double perovskite proposed to be at the border between two and three dimensional magnetism, with a square lattice of $S=\frac{1}{2}$ Cu$^{2+}$ ions. We have used inelastic neutron scattering to investigate the spin wave excitations of the system, to find out how they evolve as a function of temperature, as well as to obtain information about the magnetic exchange interactions. We observed well defined dispersive spin wave modes at $6$~K, which partially survive above the magnetic ordering temperature, $T_N=24$~K.
Linear spin wave theory is used to determine the exchange interactions revealing them to be highly two-dimensional in nature. Density functional theory calculations are presented supporting this experimental finding, which is in contrast to a previous \emph{ab-initio} study of the magnetic interactions.
Our analysis confirms that not the nearest neighbour, but the next nearest neighbour interactions in the tetragonal $ab$ plane are the strongest. Low incident energy measurements reveal the opening of a $0.6(1)$~meV gap below $T_N$, which suggests the presence of a very weak single ion anisotropy term in the form of an easy axis along $\hat{\mathbf{a}}$.
\end{abstract}


\maketitle

\section{INTRODUCTION}
Low dimensional magnetism is currently of great interest to condensed matter physics, partly due to the link to the two dimensional antiferromagnetic parent phases of the high-$T_c$ superconductors\cite{Dai,Keimer}. In those compounds it is considered that the square lattice of $S=\frac{1}{2}$ Cu$^{2+}$ $3d^9$ ions is responsible for their magnetic and superconducting behaviour. They possess strong in-plane nearest neighbour (NN) superexchange ($J\sim130$~meV) and weaker next nearest (NNN) exchange ($J'\sim18$~meV)\cite{Hayden,Coldea,Headings}. Similar copper square lattice compounds with weaker interactions are of interest as a point of comparison for the fundamental understanding of the magnetism of square lattices of Cu$^{2+}$ ions.

The $B$-site ordered double perovskite oxides Sr$_2$Cu$B''$O$_6$, where $B''$ is a diamagnetic hexavalent ion such as Mo, Te or W, are examples of such materials\cite{Vasala3}. Although they are structurally three dimensional, many display low dimensional properties. The $ab$ planes have a square centred array of Cu$^{2+}$, with the half-filled Cu $3d_{x^2-y^2}$ orbitals ordered into the $ab$ planes by the Jahn-Teller distortion, creating magnetic interactions between the neighbouring Cu ions within the $ab$ planes. As the $d_{z^2}$, $d_{yz}$ and $d_{zx}$ orbitals are all filled, the magnetic interactions along the $c$ axis are expected to be weak, resulting in the magnetic interactions being quasi two-dimensional\cite{Vasala4}. Compared to the interactions in the cuprates, in the double perovskites the magnetic ions are separated by an array of diamagnetic O-$B''$-O ions, making them an order of magnitude weaker, and therefore it is possible to study the low temperature low dimensional magnetic properties.

\begin{figure}
	\begin{center}
		\includegraphics[width=.6\linewidth, angle=0]{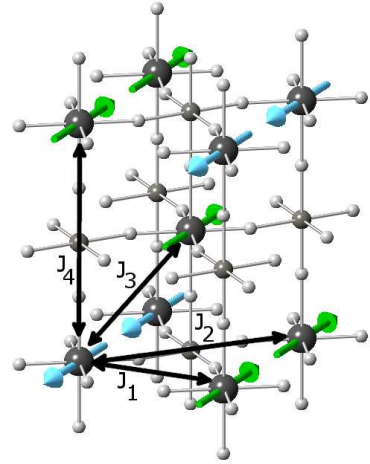}
	\end{center}
	\caption{Schematic view of the magnetic structure and exchange constants $J_1$ to $J_4$ between Cu ions in double perovskite \scwo. Large dark spheres are Cu, smaller spheres are W and the smallest palest spheres are O. Sr has been omitted for clarity.}
	\label{fig:struc}
\end{figure}

\scwo\hspace{2pt} shows a broad maximum in the magnetic susceptibility at $83$~K, behaviour characteristic of two-dimensional Quantum Square Lattice Heisenberg Antiferromagnets (QSLHAF), with no clear indication of a transition to a long-range-ordered magnetic state\cite{Vasala1}. There is a kink in the second derivative at $24$~K, but the data are inconclusive of any transition. Instead the transition to a long-ranged ordered state was confirmed definitively via $\mu$SR, with the observation of a spontaneous oscillation below $24$~K, and only a slowly decaying component above, with no sign of quasistatic short-range order\cite{Vasala1}. Regrettably there was insufficient data for the temperature dependence of the local magnetic field to determine unambiguously whether the ordering is $3D$ or lower dimensional.

High flux neutron powder diffraction data have revealed that the Cu$^{2+}$ ions display antiferromagnetic Type-II ordering with $0.57(1)\mu_\mathrm{B}$ magnetic moments aligned along the $a$-axis\cite{Vasala2}, see Figure~\ref{fig:struc}. The moment is smaller than might be expected for $S=\frac{1}{2}$ Cu$^{2+}$ ions ($m_s=g\times j=g\times s=1\mu_\mathrm{B}$), possibly due to a degree of frustration or being on the borderline between quasi-low dimensional and three dimensional magnetism. It may also be the signature of quantum zero-point fluctuations, which can reduce the moment in QSLHAF systems\cite{Reger}. 

Electronic structure calculations, using a Coulomb $U$ value determined in relation to oxygen $K$-edge X-ray absorption spectroscopy measurements, gave the exchange constants: $J_1=-1.20$, $J_2=-7.47$, $J_3=-0.03$ and $J_4=-4.21$~meV, resulting in a reasonable agreement with the measured Curie-Weiss temperature ($\theta_\mathrm{meas}=-116$~K, $\theta_\mathrm{calc}=-126$~K)\cite{Vasala1}. These values are consistent with the observed Type II antiferromagnetic structure\cite{Vasala2}, but the interplanar $J_4$ coupling indicates significantly stronger three dimensional magnetism than might have been expected from the electronic structure and based on the form of the magnetic susceptibility. To investigate this apparent discrepancy we have performed inelastic neutron scattering (INS) measurements and a comprehensive Density Functional Theory (DFT) study to re-examine the exchange constants in \scwo.

In this paper we present our inelastic neutron scattering measurements performed on \scwo. The INS results are analysed and compared with linear spin wave theory simulations based on the original DFT estimates\cite{Vasala1} for the exchange interactions. This demonstrates a significant disagreement with the earlier DFT calculations. We have reassessed the DFT calculations and present revised results, which support the conclusions drawn from our INS data indicating a strong two dimensional character, and revealing the significance of the straight Cu-O-W-O-Cu linkers.

\section{EXPERIMENTAL DETAILS}

A $6.89$~g powder sample of \scwo was synthesized by solid-state reaction of a stoichiometric mixture of SrCO$_3$, CuO and WO$_3$ powders, according to the method detailed in Ref.~\onlinecite{Vasala1}. The phase purity and quality of the sample were verified using x-ray diffraction (X'Pert Pro MPD, Cu $K_{\alpha_{1}}$ radiation). Rietveld refinement using the FULLPROF program\cite{FullProf} confirmed that the sample is single phase with the $I4/m$ structure and lattice parameters $a=5.430(2)$~\AA\ and $c=8.415(2)$~\AA\ as reported earlier\cite{Vasala1}.

Neutron inelastic scattering measurements were performed on the MERLIN time-of-flight direct geometry spectrometer\cite{Bewley} at the ISIS facility of the Rutherford Appleton Laboratory. The sample was contained in an aluminium foil packet in the form of an annulus of diameter $40$~mm and height $40$~mm and sealed in a thin aluminium can containing helium exchange gas. The sample can was cooled by a closed-cycle refrigerator. The straight Gd slit package was used in the Fermi chopper, which was phased to allow the recording of spectra with incident energies of either $18$ and $45$~meV (at a rotation speed of $250$~Hz), or $10$ and $34$~meV (at a rotations speed of $150$~Hz) simultaneously via the rep-rate multiplication method\cite{Russina09,Russina10,Nakamura}. The data were collected at a series of temperatures between $T=6$ and $93$~K for $\sim4$~h each. The data were reduced using the MantidPlot software package\cite{Arnold}. The raw data were corrected for detector efficiency and time independent background following standard procedures\cite{Windsor}. Vanadium spectra were recorded\cite{noteVan} with the same incident energies to determine the energy resolution and to convert the intensities into units of cross section, mb.sr$^{-1}$.meV$^{-1}$.f.u.$^{-1}$, where f.u. stands for the formula unit of \scwo. Additional measurements with $E_i=1.9, 3.5, 8.3$~meV at a chopper frequency of $100$~Hz were performed on the same sample on the LET time-of-flight direct geometry spectrometer\cite{BewleyLET} also using the rep-rate multiplication method.

\begin{figure*}
    \begin{center}
        \includegraphics[width=.85\linewidth]{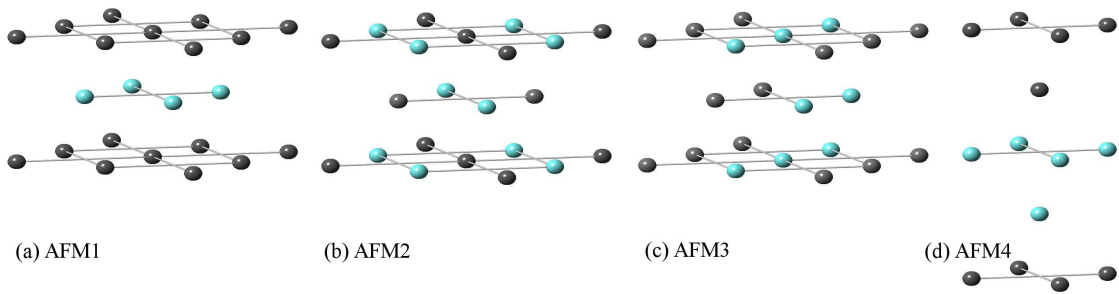}
    \end{center}
    \caption{\label{fig:AFs}Schematic of the four AFM orderings used in calculating the magnetic exchange constants of \scwo. Black and cyan spheres correspond to the two spin states on the Cu ions. Other ions and long bonds along the $c$-axis are omitted for clarity.}
\end{figure*}

\section{COMPUTATIONAL DETAILS}
The exchange constants $J_1-J_4$ can be obtained using DFT by calculating the energy differences between multiple collinear spin states and projecting those onto the following Hamiltonian:
\begin{equation}\label{eq1}
H=-\sum_{ij}J_{ij}\mathbf{S}_i\cdot\mathbf{S}_j,
\end{equation}
This is known as the mapping method.\cite{Koo1,Koo2} One ferromagnetic and four antiferromagnetic collinear spin states are sufficient in \scwo, see ref.\onlinecite{Vasala1} for more details. These configurations are presented in Figure~\ref{fig:AFs} and consist of $2\times2\times1$ (AFM1-3) or $1\times1\times2$ (AFM4) supercells. The exchange constants can be solved from the following equations:
\begin{equation}\label{eq2}
J_3=\left(E_\mathrm{AFM1}-E_\mathrm{FM}\right)/16S^2,
\end{equation}
\begin{equation}\label{eq3}
J_1=\left(E_\mathrm{AFM2}-E_\mathrm{FM}-8J_3S^2\right)/8S^2,
\end{equation}
\begin{equation}\label{eq4}
J_2=\left(E_\mathrm{AFM3}-E_\mathrm{FM}-4J_1S^2-8J_3S^2\right)/8S^2,
\end{equation}
\begin{equation}\label{eq5}
J_4=\left(E_\mathrm{AFM4}-E_\mathrm{FM}-8J_3S^2\right)/4S^2.
\end{equation}

Total energies of the spin configurations were determined by means of density functional theory calculations using the full-potential linearised augmented plane-wave plus local orbitals (FP-LAPW+lo) ELK code\cite{ELK}. The calculations were performed using the experimental crystal structure of \scwo\ determined by neutron diffraction\cite{Vasala1}. The generalised gradient approximation (GGA) exchange and correlation functionals by Perdew, Burke and Ernzerhof were used\cite{Perdew}. A $k$ point mesh of either $4\times4\times6$ or $8\times8\times3$ was used depending on the supercell. The plane-wave cutoff was set at $\left|G+k\right|_\mathbf{max}=8/R_{MT}$~a.u.$^{-1}$, where $R_{MT}$ is the radius of the smallest muffin-tin (oxygen, $1.55$~a.u.).

\scwo\ is a strongly correlated material, and thus electron correlation effects are central for modelling the electronic structure. The correlation effects of localised Cu$^{2+}$ $3d$ electrons were included within the semi-empirical DFT+$U$ framework with Hubbard $U$ and Stoner $I$ as parameters\cite{Anisimov}. DFT+$U$ methods such as GGA+$U$ require the use of a double counting correction, since Coulomb and intra-atomic exchange interactions are also included in the GGA functionals. We have used both Around Mean Field (AMF)\cite{Czyzyk} and Fully Localised Limit (FLL)\cite{Liechtenstein} double counting corrections in this work. Exchange constants calculated by DFT are known to be sensitive to the on-site Coulomb term $U$ and the double counting correction used\cite{Lebernegg}. For this reason, we have calculated the exchange constants using a range of $U$ values typical for Cu $3d$ with two different double counting corrections. With the FLL correction a Hubbard $U$ of $\sim8-9$~eV has been widely used for Cu in oxides,\cite{Anisimov,Liechtenstein,Lebernegg,Chou} whereas a slightly lower $U$ of $\sim6-7$~eV is typical when using the AMF correction\cite{Czyzyk,Lebernegg}. The intra-atomic exchange parameter $I$ was chosen to be $0.9$~eV, which is a common value in $3d$ transition metal oxides\cite{Anisimov}. 

In order to evaluate the overall strength of the exchange interactions with different Hubbard $U$ values, we have calculated the Weiss temperature $\theta$ using the mean field approximation:
\begin{equation}
\theta=\frac{S(S+1)}{3k_B}\sum_i z_iJ_i,
\end{equation}
where $k_B$ is the Boltzmann constant and $z_i$ is the number of sites connected by exchange interaction $J_i$.

\section{Experimental Results}

\begin{figure}[b]
    \begin{center}
        \includegraphics[width=.75\linewidth]{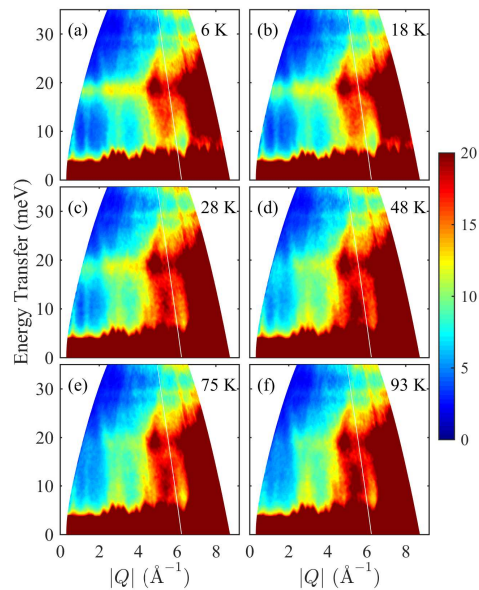}
    \end{center}
    \caption{\label{fig:colplot}(Colour online) Colour coded inelastic neutron scattering intensity maps in units of mb.sr$^{-1}$.meV$^{-1}$.f.u.$^{-1}$, energy transfer vs momentum transfer $Q$ of \scwo\hspace{1pt} measured with an incident energy of $E_i=45$~meV on MERLIN.}
\end{figure}

\begin{figure*}
    \begin{center}
        \includegraphics[width=.75\linewidth, angle=0]{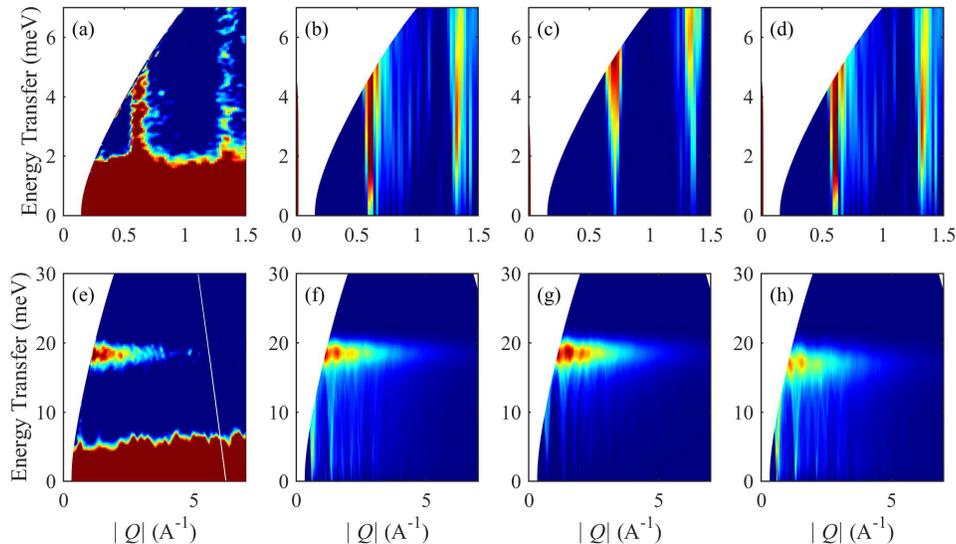}
    \end{center}
    \caption{\label{fig:phonsub}(Colour online) The measured magnetic scattering at $6$~K after phonon scattering subtraction for (a) $E_i=10$ and (e) $E_i=45$~meV; compared with the spin wave scattering at $6$~K simulated using the SpinW program with exchange parameters (b) and (f) optimised for the experimental data: $J_1=-1.2$~meV, $J_2=-9.5$~meV, $J_3=0$~meV and $J_4=-0.01$~meV, (c) and (g) from a previous DFT study\cite{Vasala1}: $J_1=-1.20$~meV, $J_2=-7.47$~meV, $J_3=-0.03$~meV and $J_4=-4.21$~meV, and (d) and (h) from the DFT study presented in Section~V: $J_1=-2.45$~meV, $J_2=-8.83$~meV, $J_3=0$~meV and $J_4=-0.01$~meV.}

\end{figure*}

For temperatures below $T_N$, additional peaks are observed at the elastic line at $|Q|=0.69$~\AA\ and $1.35$~\AA, corresponding to the ($0$ $\frac{1}{2}$ $\frac{1}{2}$) and ($1$ $\frac{1}{2}$ $\frac{1}{2}$) magnetic Bragg peaks, confirming the previous assignment of Type II antiferromagnetic ordering described by a [$0$ $\frac{1}{2}$ $\frac{1}{2}$] magnetic ordering wavevector\cite{Vasala2}. The colour-coded inelastic neutron scattering intensity maps of \scwo\hspace{1pt} measured on MERLIN at various temperatures between $T=6$ and $93$~K are shown in Figure~\ref{fig:colplot}(a-f).
At low temperatures, for momentum transfer $|Q|<4$~\AA$^{-1}$, a strong flat scattering band can be observed at $\sim18$~meV (Fig.~\ref{fig:colplot}). In addition steep spin waves are seen apparently emanating from the elastic line. Based on the measurements performed with an incident energy of $10$~meV on MERLIN, these excitations would appear to be gapless to within the resolution of the instrument (FWHM$=0.65\pm0.01$~meV). At larger $|Q|$ values the excitations are dominated by phonons.

Looking at the temperature evolution, the flat band at $\sim18$~meV appears to be more strongly effected by increasing temperature, disappearing between $28$~K and $48$~K, while evidence of the spin waves persists up to at least $75$~K, i.e. well above $T_N=24$~K, which indicates the presence of two dimensional interactions. These features are both absent at $93$~K, above the broad maximum seen in the magnetic susceptibility at $T_\mathrm{max}=83$~K\cite{Vasala1}. The assignment of the higher $|Q|$ features as phonons is further confirmed by their increasing intensity with increasing temperature. By considering the Bose factor, and using the $93$~K data we can subtract the phonons from the low temperature data to give the purely magnetic signal\cite{Carlo}, as shown in Fig.~\ref{fig:phonsub}(a) and (e).

In order to model the observed magnetic spectrum, we have calculated the spin wave dispersions, the spin-spin correlation function and the neutron scattering cross section using the SpinW program\cite{SPINW}. Since tungsten is hexavalent in \scwo, it is diamagnetic, and therefore only interactions between the Cu$^{2+}$ ions need to be considered. We have constructed the magnetic Hamiltonian in Eq.~\eqref{eq1} with four different exchange couplings for the nearest and next nearest neighbour interactions in plane, and interplanar along the c-axis, as shown in Fig.~\ref{fig:struc}.
As we could not confirm the existence of a small spin gap to within the instrumental resolution of MERLIN, we initially neglect a single ion anisotropy term, which would generally open such a gap.
For the spin wave calculation we started from the original DFT-calculated values of the exchange parameters given in Section I. Further the instrument resolution was estimated from vanadium runs and this was included in the simulation.

\begin{figure}
    \begin{center}
        \includegraphics[width=.75\linewidth, angle=0]{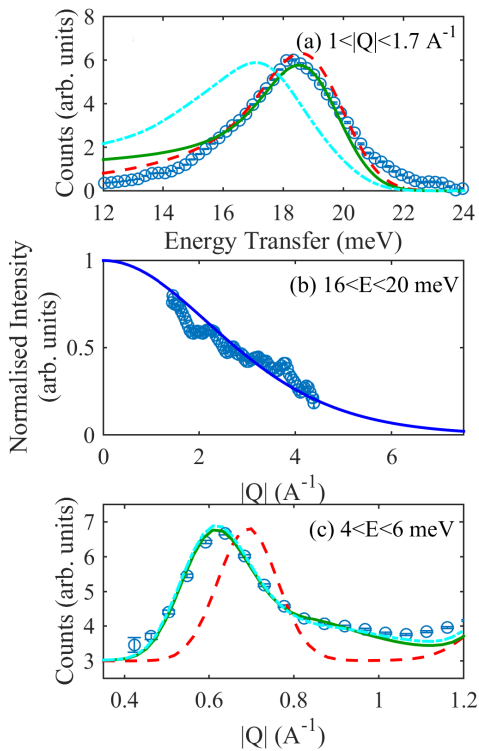}
    \end{center}
    \caption{\label{fig:cuts}(Colour online) One dimensional cuts of the magnetic scattering from \scwo\hspace{1pt} at $6$~K (a) intergrated between $1<|Q|<1.7$~\AA$^{-1}$ compared with the simulated powder average spin waves for the original DFT-determined exchange parameters (dashed red line) and for the more two dimensional INS obtained exchange parameters (solid green line), and for the best revised DFT exchange parameters (dot-dash cyan line), (b) a q-cut through the $\sim18$~meV feature, showing that it follows the magnetic form factor for Cu$^{2+}$, and (c) a q-cut through the data integrated between $4<E<6$~meV compared with simulations for the original DFT-determined exchange parameters (dashed red line), for the more two dimensional INS obtained exchange parameters (solid green line) and for the best revised DFT exchange parameters (dot-dash cyan line).} 
\end{figure}

Comparing the simulation for the original DFT-determined exchange parameters\cite{Vasala1} (Figs.~\ref{fig:phonsub}(c) \& (g)) with the measured spin wave dispersion (Figs.~\ref{fig:phonsub}(a) \& (e)) we see that the simulation accurately reproduces the band maximum at $18$~meV, with steep spin waves emerging from the elastic line. However, closer inspection of the data reveals that the spin waves do not actually emerge from the magnetic Bragg peaks. Instead the centre of the lower $|Q|$ excitation is at $0.62$~\AA$^{-1}$, lying between $0.58$~\AA$^{-1}$, corresponding to the forbidden $(\frac{1}{2} 0 0)$ position, and $0.69$~\AA$^{-1}$, corresponding to the allowed $(\frac{1}{2} 0 \frac{1}{2})$ position. In order to better reproduce the data, we have found a new set of exchange parameters: $J_1=-1.2$~meV, $J_2=-9.5$~meV, $J_3=0$~meV and $J_4=-0.01$~meV, which are noticeably more two dimensional than those obtained originally using DFT\cite{Vasala1}, see the simulation panels Fig.~\ref{fig:phonsub}(b) and (f).
The difference in the two simulations can be seen more strikingly in Figure~\ref{fig:cuts}. While both manage to reproduce the band maximum at $\sim18$~meV (Fig.~\ref{fig:cuts}(a)), which is shown to follow the magnetic form factor for Cu$^{2+}$ in Fig.~\ref{fig:cuts}(b); the cut through the excitations integrated for $4<E<6$~meV (Fig.~\ref{fig:cuts}(c)), shows how the original DFT-determined exchange parameters simulation inaccurately estimates the $|Q|$ position of the first excitation as emerging from the ($0$ $\frac{1}{2}$ $\frac{1}{2}$) magnetic Bragg peak, while the simulation for the new exchange parameters reproduces the data well. When the simulation for the new experimental exchange parameters is performed to obtain the single crystal dispersion, it becomes clear that the excitations are emerging from both the $(\frac{1}{2} 0 \frac{1}{2})$ and $(\frac{1}{2} 0 0)$ positions, and it is the powder averaging that gives rise to the observed $0.62$~\AA$^{-1}$ position. In order to understand the appearance of the softening at the forbidden $(\frac{1}{2} 0 0)$ position, it is necessary to look at the value of $J_4$. If $J_4$ were ferromagnetic, then Type I antiferromagnetic order would be stabilised. Our value for $J_4$ of only $-10\mu$eV puts \scwo\ close to the border between Type I and Type II ordering, such that softening is observed at both positions, but the negative sign results in Type II order.

In optimising our exchange parameters to fit the data, it is made clear that the value of $J_2$ is central to the position of the band maximum. Simulations for the spin wave dispersion indicate that the value of $J_1$ modulates the spin wave maxima over different positions in the Brillouin Zone, that on powder averaging leads to a broadening of the $18$~meV band maximum. We have estimated $J_1$ based on the width of the peak in Fig.~\ref{fig:cuts}(a), taking the instrumental resolution at an energy transfer of $18$~meV into account.

\begin{figure}
    \begin{center}
        \includegraphics[width=1\linewidth, angle=0]{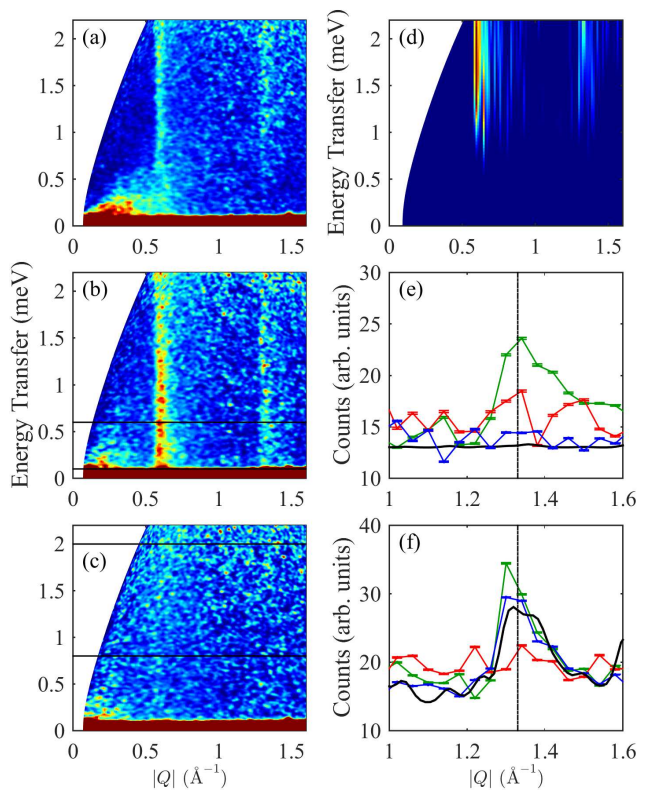}
    \end{center}
    \caption{\label{fig:LET}(Colour online) The magnetic scattering data for $E_i=3.51$~meV at (a) $5$~K, (b) $30$~K and (c) $100$~K. (d) shows a SpinW simulation for the low temperature data, reproducing a gap with the addition of single ion anisotropy term. One dimensional cuts through the data at $5$~K (blue), $30$~K (green) and $100$~K (red) for (e) $0.1<E<0.6$~meV and (f) $0.8<E<2$~meV are compared with cuts through the simulation (black line).}
\end{figure}

\begin{table*}
    \caption{\label{Tab:DFT} Exchange constants of \scwo\ calculated by GGA+$U$ using the mapping method compared with the results obtained from inelastic neutron scattering.}
    \begin{tabular}{l|>{\hfill}p{1cm}>{\hfill}p{1cm}>{\hfill}p{1cm}>{\hfill}p{1cm}|>{\hfill}p{1cm}>{\hfill}p{1cm}>{\hfill}p{1cm}|>{\hfill}p{1cm}}
        \hline\hline
        &   \multicolumn{4}{c|}{AMF} &   \multicolumn{3}{c|}{FLL} & \multicolumn{1}{c}{INS}\\
        \hline
        $U$ (eV)        & 5         & 6         & 7         & 8         & 7     & 8     & 9     & -\\
        $J_1$ (meV)     & -3.70     & -3.18     & -2.70     & -2.56     & -3.29 & -2.45 & -2.58 & -1.2\\
        $J_2$ (meV)     & -13.83    & -10.87    & -7.96     & -6.23     & -10.42& -8.83 & -6.75 & -9.5\\
        $J_3$ (meV)     & 0.01      & 0.06      & -0.06     & -0.04     & -0.02 & 0.03  & -0.04 & 0\\
        $J_4$ (meV)     & 0.12      & 0.03      & 0.19      & 0.07      & 0.09  & 0.04  & -0.12 & -0.01\\
        $J_2/J_1$       & 3.73      & 3.42      & 2.95      & 3.61      & 3.16  & 3.61  & 2.61  & 7.92\\
        $\theta$ (K)    & -202.5    & -161.4    & -102.4    & -102.4    & -159.1& -129.9& -110.0& -124.1\\
        \hline
    \end{tabular}
\end{table*}

While the energy resolution on MERLIN did not allow us to observe a spin gap, by using the cold chopper spectrometer LET we are able to distinguish the presence of a small gap of order $0.6(1)$~meV below $T_N$. Figure~\ref{fig:LET} shows the temperature evolution of this gap. Below $T_N$ (Fig.~\ref{fig:LET}(a)) the gap is present, although somewhat obscured for the lower $|Q|$ excitation by the presence of a spurious signal close to the beam stop. When the temperature is raised above $T_N$ the gap closes, with strong spectral weight shifted down to the elastic line (Fig.~\ref{fig:LET}(b)). Finally, above $T_\mathrm{max}$, the excitations are entirely absent. We have been able to reproduce the presence of the gap in the ordered structure in our SpinW simulations (Fig.~\ref{fig:LET}(d)) by modifying Eq.~\eqref{eq1} to include single ion anisotropy in the form of an easy axis along $\hat{\mathbf{a}}$ of magnitude $0.025$~meV. Although it is commonly believed that magnetic systems of spin-$1/2$ transition-metal ions have no magnetic anisotropy arising from spin-orbit coupling\cite{Moriya}, this is only the case for ions in a perfect octahedral crystal field ($t_{2g}$ ground state). Any distortion which splits the triply degenerate ground state and mixes the orbital d-states can, via the spin-orbit coupling, give non-zero single ion anisotropy\cite{Liu}. In Fig.~\ref{fig:LET}(e) a cut along $|Q|$ integrating between $0.1$ and $0.6$~meV shows how peaks are only observed in the data for $T_N<T<T_\mathrm{max}$, while in Fig.~\ref{fig:LET}(f) a cut integrating energies between $0.8$ and $2.0$~meV shows peaks for both $T_N<T<T_\mathrm{max}$ and $T<T_N$, where the lowest temperature data is well matched by the simulation.

\section{Computational Results}

The exchange constants calculated by DFT are presented in Table~\ref{Tab:DFT}. In all cases the main interactions are the in-plane $J_1$ and $J_2$ interactions, which are antiferromagnetic. These results show that the next-nearest neighbor $J_2$ interaction in \scwo\ is stronger than the nearest neighbor $J_1$ interaction, which is consistent with the experimental Type II magnetic structure\cite{Vasala2}. The strength of the exchange interactions decreases with increasing Hubbard $U$, which is clearly seen from the change in Weiss temperatures. This decrease is typical when the mapping method is used with DFT+$U$\cite{Koo1,Koo2}.

The exchange constants obtained by AMF and FLL double counting corrections are similar, with FLL producing stronger interactions for the same Hubbard $U$. The computational results are in fairly good, but not perfect, agreement with the experimental INS results. The DFT calculations presented here consistently over-estimate the nearest-neighbour $J_1$ interaction, which results in a rather lower $J_2/J_1$ ratio ($\sim3$) than experimentally observed ($\sim8$). $J_2$, in contrast, is in good agreement with the experimental results. The inter-planar interactions $J_3$ and $J_4$ are very weak compared to $J_1$ and $J_2$ in all solutions, which is consistent with the experimental exchange constants obtained by INS. This shows that the magnetic interactions in \scwo\ are highly two-dimensional. However, the correct negative sign (i.e. antiferromagnetic interaction) for $J_4$ is only obtained with FLL correction and $U = 9$~eV. This solution also results in a Weiss temperature of $110$~K, which is the closest to the experimental value of $116$~K\cite{Vasala1}, but the low value of $J_2$ leads to a sizeable underestimate of the band maximum. If $J_3$ and $J_4$ are constrained to be small and negative, then the best match to the experimental data is obtained for the FLL double counting correction with $U=8$~eV, see Fig.~\ref{fig:phonsub}(d) and (h), although they do not exactly reproduce the inelastic neutron scattering data, underestimating the band maximum position and overestimating the width (Fig.~\ref{fig:cuts}(a)).

The previous DFT study of \scwo\cite{Vasala1} found a far stronger $J_4$ interaction of $-4.21$~meV, indicating significantly more three-dimensional magnetism than reported here. The difference between the results presented in Table~\ref{Tab:DFT} and those published previously\cite{Vasala1} is perhaps due to the new calculations being performed over a range of $U$ values with a more accurate crystal structure, higher plane-wave cut-off and a newer branch of the ELK code, although the exact origin of this discrepancy is not known.

\section{Discussion}
Now we compare the exchange parameters estimated for \scwo\ with those of Sr$_2$CuTeO$_6$\cite{Koga,Babkevich}. Based on magnetic susceptibility measurements, it is estimated that in isostructural Sr$_2$CuTeO$_6$ the ratio $J_2/J_1<0.07$ \cite{Koga}, compared to $\sim8$ in \scwo\ from the INS results, while recent inelastic neutron scattering measurements have reduced the Sr$_2$CuTeO$_6$ value even further to $0.03$ \cite{Babkevich}. Koga \emph{et al.} propose that this reversal in relative strengths of nearest and next nearest neighbour interactions is due to the outermost filled orbital in Te$^{6+}$ being $d_{x^2-y^2}$, such that the two hole spins must be antiparallel, giving antiferromagnetic exchange $J_1$ and $J_2$; whereas in W$^{6+}$ the $p_x$ and $p_y$ orbitals are orthogonal so that the two hole spins are parallel, which would give a ferromagnetic exchange for $J_1$ and an antiferromagnetic exchange for $J_2$. However, our DFT calculations and spin wave analysis indicates a weak but antiferromagnetic $J_1$ in \scwo, which might be due to the presence of frustration. Babkevich \emph{et al.} have combined their INS measurements on Sr$_2$CuTeO$_6$ with \emph{ab-initio} calculations, which have revealed that, in fact, the dominant exchange path is via Cu-O-O-Cu, and not via the Te $4d$ orbitals.

It is interesting to also compare the inelastic neutron scattering results from $3d$ double perovskite \scwo\hspace{1pt} with those reported for $4d$ and $5d$ double perovskites with monoclinic, tetragonal or cubic crystal structures. It is to be noted that in the cubic (or less distorted monoclinic) structure of double perovskites with a single magnetic ion, the magnetic lattice is face centred cubic, which is a geometrically frustrated lattice and provides a unique opportunity to investigate frustrated magnetism. The inelastic neutron scattering study of face centred cubic Ba$_2$YMoO$_6$ (Mo$^{5+}$ $4d^1$ $S=\frac{1}{2}$) demonstrates the existence and temperature dependence of a gapped magnetic excitation at $28$~meV, with a bandwidth of $4$~meV\cite{Carlo}. The observed dispersive triplet excitations come from a singlet ground state formed from orthogonal dimers on the Mo$^{5+}$ tetrahedra. On the other hand, an inelastic neutron scattering study on monoclinic La$_2$NaRuO$_6$ also reveals a spin gap of $2.75$~meV. As the magnetic anisotropy is expected to be small for octahedrally-coordinated Ru$^{5+}$ $4d^3$ $S=\frac{3}{2}$ systems, the large gap observed for La$_2$NaRuO$_6$ may originate from the significantly enhanced value of the spin-orbit coupling in this $4d$ material\cite{Aczel}. FCC Ba$_2$YRuO$_6$ also displays a $\sim5$~meV spin gap, with a zone boundary energy of $14$~meV, at the $[100]$ magnetic ordering wavevector below $T_N=26$~K\cite{Carlo2}. INS has also revealed well defined dispersive spin wave excitations in a polycrystalline sample of monoclinic Sr$_2$YRuO$_6$, with a zone boundary energy of $\sim8$~meV at $T=5$~K and a gap of $1.2$~meV below $20$~K, but gapless above, despite being well below $T_N=31$~K \cite{Adroja}. The presence of strong diffusive scattering between $T_N$ and $300$~K is indicative of strong magnetic frustration between Ru-Ru atoms. The estimated exchange interactions give a ratio between nearest and next nearest neighbours $J_2/J_1\sim0.14$, revealing much stronger nearest neighbour interactions in contrast to \scwo ($J_2/J_1\sim8$).

Further instructive comparison might be made with other $S=1/2$ Cu$^{2+}$ quantum square lattice Heisenberg antiferromagnets (QSLHAF), which have been of considerable interest both theoretically and experimentally ever since the realisation that the parent compounds of the cuprate superconductors could be described using the same model. The low energy dynamics of QSLHAF are well described using linear spin wave theory with quantum corrections. However, inelastic neutron scattering measurements on a range of Cu$^{2+}$ QSLHAF have revealed a glaring anomaly at high energy in the vicinity of $\mathbf{q}=(\pi,0)$, where the intensity of the otherwise sharp excitations is completely wiped out\cite{Plumb,Headings,Tsyrulin,Christensen,Dalla}. Identifying the origin of this effect is complicated by the presence of additional exchange terms such as electronic ring exchange\cite{Plumb,Headings} and further neighbour exchange\cite{Tsyrulin}, as is also present in \scwo. Due to similarities in the measured anomaly with predictions for fermionic Resonating Valence Bond excitations\cite{Ho}, it has been speculated that the anomaly may be related to fractionalised spin excitations\cite{Headings,Christensen}. By analogy with 1-D systems, these are referred to as spinons. In 1-D spinons have been identified in a number of materials, but observing 2-D analogues has proved more challenging until recently\cite{Dalla}. It would therefore clearly be very interesting to measure the excitations in single crystal \scwo.

\section{CONCLUSIONS}
We have performed inelastic neutron scattering measurements on double perovskite \scwo, which reveal clear evidence of spin wave excitations at low temperatures. The magnetic excitations partially survive at temperatures above $T_N=24$~K for long range 3D order, indicating a 2D component to the nature of the magnetic interaction. Our spin wave analysis using linear spin wave theory indicates that the NNN interaction in the $ab$ plane is a factor of approximately eight times stronger compared to the NN interaction in the $ab$ plane. While a previous DFT study gave a ratio of $1.78$ for the strong interactions in plane ($J_2$) and interplane ($J_4$), our inelastic neutron scattering results indicate that $J_2$ is significantly stronger than $J_4$, which is consistent with the expected two dimensional behaviour given the Jahn-Teller distortion. The more comprehensive DFT study presented here has obtained results supporting this two dimensional nature and the dominance of the NNN interaction. The strongest interaction in the $ab$ plane is most probably arising due to superexchange between the Cu$^{2+}$ $d_{x^2-y^2}$ orbitals via the Oxygen $p$ orbitals along straight linkers. Furthermore, the observation of a very small spin gap in \scwo\hspace{1pt} is in line with a general explanation, which attributes the opening of increasingly large spin gaps in $4d$ and $5d$ systems as being due to the stronger spin orbit coupling compared to that in $3d$ systems.

\section*{}
Upon submission we were made aware of another paper reporting inelastic neutron scattering measurements on \scwo\cite{Burrows}.

\begin{acknowledgements}
The authors acknowledge CSC-IT Centre for Science, Finland, for providing computational resources. DTA would like to thank JSPS for funding his visit to Hiroshima University.
\end{acknowledgements}

\bibliography{scwo}

\begin{thebibliography}{43}
\bibitem [{not()}]{noteVan}%
  \BibitemOpen
  \href@noop {} {}\bibinfo {note} {These were performed on a $7.85$~g
  cylindrical vanadium ($99.8$~\% purity) reference sample.}\BibitemShut
  {Stop}%
\end{thebibliography}

\end{document}